\documentclass[twocolumn,showpacs,amsmath,amssymb, showkeys]{revtex4}

\usepackage{graphicx}
\usepackage{dcolumn}
\usepackage{bm}

\begin{document}

\title{Study of low energy hadronic interaction models based on BESS observed cosmic ray proton and antiproton spectra at medium high altitude}

\author {Arunava Bhadra$^{1}$, Sanjay K. Ghosh$^{2,3}$, Partha S. Joarder$^{3}$, Arindam Mukherjee$^{1}$, and Sibaji Raha$^{2,3}$}

\affiliation{$^{1}$ High Energy $\&$ Cosmic Ray Research Centre, University of North Bengal, Siliguri, WB 734013 India \\
$^{2}$ Department of Physics, Bose Institute, 93/1 A.P.C. Road, Kolkata, India 700009 \\
$^{3}$ Centre for Astroparticle Physics and Space Science, Bose Institute, Block EN,Sector V, Salt Lake, Kolkata, India 700091}

\begin{abstract}
We study low energy hadronic interaction models based on BESS observed cosmic ray proton and antiproton spectra at medium high altitude. Among the three popular low energy interaction models, we find that FLUKA reproduces results of BESS observations on secondary proton spectrum reasonably well over the whole observed energy range, the model UrQMD works well at relatively higher energies whereas spectrum obtained with GHEISHA differs significantly from the measured spectrum. Simulated antiproton spectrum with FLUKA, however, exhibits significant deviations from the BESS observation wheras UrQMD and GHEISHA reproduce the BESS observations within the experimental error. 
\end{abstract}

\maketitle

\section{Introduction}
Precise examination of the development of cosmic ray shower in the earth's atmosphere is important in various contexts such as in the study of atmospheric neutrino oscillations or in the study of ultra high energy cosmic rays. The detailed process of development of such an extensive air shower (EAS) is, however, too complicated to be amenable to simple analytical descriptions. The main results concerning the flux and the other important features of the secondary cosmic ray particles in an EAS are thus obtained principally through the Monte Carlo (MC) simulation techniques. 

MC simulations of the extensive air shower is strongly dependent on our knowledge of the interaction mechanisms of energetic particles. Such knowledge on particle interactions is somewhat uncertain at high energies as the accelerator data for relevant target-projectile combinations covering the whole kinematic region are not yet available. Even at low (below $\sim 5$ GeV) and intermediate (from few GeV to few hundred GeV) energies, there is a lack of data on hadron-nucleus interactions and almost no measurements are available for the particle production in pion-nucleus collisions. One, therefore, relies mostly on theoretical models of particle interactions in such cases. The interaction models used in different simulation programs are necessarily of the nature of extrapolations of the known processes and/or of the low energy accelerator data so that each of such models has its own parameterization guided by some (mainly QCD-motivated) theoretical prescriptions. The limited knowledge of particle interactions is considered to be one of the main sources of uncertainty in the estimation of the secondary particle flux in an EAS. 

In view of the large uncertainties involved in the description of high energy particle interactions, the influence of high energy hadronic models on  air shower observables has been a topic of active reserach for quite some time. Recent studies, however, suggest that the low and the intermediate energy hadronic interaction models also play crucial role in the precise estimation of the low energy secondary cosmic ray flux in the atmosphere [Wentz et al., 2003; Djemil, Attallah and Capdevielle, 2007; Sanuki et al., 2007] simultaneously influencing some of the important characteristics of the extensive air showers. A strong dependence of the lateral particle distribution of the simulated extensive air showers at large core distances on the intermediate energy hadronic interaction models has been reported [Drescher et al., 2004].

The aim of the present work is to examine the sensitivity of the low energy secondary proton and antiproton fluxes on the hadronic interaction models that are particularly suitable for use in the intermediate energy range. The simulated showers of secondary protons, that arise from hadronic interactions of the forward kinematic region, are expected to be  particularly responsive to the choice of interaction models. To generate secondary fluxes for different models of hadronic interactions, the air shower simulation program CORSIKA(COsmic Ray SImulation for KAscade) version 6.600 [Heck et al., 1998; www-ik.fzk.de/˜heck/corsika/] is exploited here. A novel feature of the CORSIKA  program is that it allows one to choose any of the three popular models, namely GHEISHA [Fesefeldt, 1985], FLUKA [Fass`o et al., 2000
; www.fluka.org/] and UrQMD [Bleicher et al., 1999; www.th.physik.uni-frankfurt.de/\~ urqmd/], for the portrayal of the intermediate energy hadronic interactions as well as one of the seven different models, namely DPMJET [Ranft, 1995], HDPM [Capdevielle et al., 1992], QGSJET 01 [Kalmykov, Ostapchenko and Pavlov, 1997], SIBYLL [Fletcher et al., 1994; Engel et al., 1992], VENUS [Werner, 1993], NEXUS [Drescher et al., 2001; www-subatech.in2p3.fr/\~theo/nexus] and QGSJET II [Ostapchenko, 2005], for the description of hadronic interactions at high energies. The borderline between high and low energies is set as $80$ GeV/n by default in this simulation program. Whether the intermediate energy interaction models of CORSIKA can be discriminated from actual measurements will also be examined in this paper by comparing the simulated secondary proton and antiproton spectrum obtained by using different low/intermediate energy interaction models with the observations.  

A practical problem in differentiating the influence of hadronic interaction models is that the estimation of secondary fluxes through the MC simulations include various systematic errors caused not only by the built-in uncertainties in various interaction models but also by the errors involved in the estimation of values of other physical inputs, such as the primary cosmic ray fluxes, the atmospheric density profiles etc. The effects of such errors in the calculation of physical inputs considerably complicate the simulated flux in such a way so as to make it difficult to isolate out the influence of the interaction models alone on the calculated values of the secondary fluxes. In particular, a dominant systematic error in evaluating the flux of cosmic ray secondaries arises from the uncertainities involved in the estimation of the input fluxes of the primary cosmic rays. To minimize the effects of such uncertainties, the simulated results of secondary cosmic ray protons and antiprotons would be compared here with the recent precise measurements of such fluxes at the mountain altitude ($742 g cm^{-2}$) by the BESS spectrometer [Sanuki et al., 2003]. Also, the primary cosmic ray energy spectra, as measured by the same BESS instruments [Sanuki et al., 2000], would be considered in this paper as the inputs of the simulations after taking into account the effects of solar modulation that are appropriate for the specific period of the measurement of secondary fluxes by the BESS experiment. 

The article is organized as follows: in the next section, we would provide a short account of the intermediate energy interaction models to be considered. The simulation procedure adapted in this work would be described in section 3. After a brief review of the precision measurements of hadronic and muon fluxes by the BESS spectrometer, the results of the simulations are compared with the corresponding observations in section 4. Concluding remarks are given in section 5. 

\section{Hadronic Interaction models at the intermediate energy range}

GHEISHA, UrQMD and FLUKA are among the most popular hadronic interaction models in the intermediate energy range that would hereafter be referred as the `low energy models' in accordance with the terminology used in air shower simulations. These models find ample applications in various branches of physics and Bio-physics that include the simulations of  accelerator-based experiments, detector design, simulations of the cosmic rays, neutrino physics, radiotherapy, accelerator driven systems etc. We have used FLUKA version 2006.3b,  GHEISHA version 2002d and UrQMD vesrion 1.3 in the present work. 

GHEISHA [Fesefeldt, 1985] is based on the parameterizations of accelerator data. It was originally developed as a event generator in which the energy, momentum, charge or the other quantum numbers are conserved only in the sense of an average instead of being conserved on an event-by-event basis. GHEISHA has been successfully used in the detector Monte Carlo code GEANT over the last twenty years and is currently being used as the default low energy interaction model in the air shower simulation program CORSIKA. Recent comparisons with experimental data, however, exhibits a few shortcomings of the older version of GHEISHA [Ferrari and Sala, 1996 ; Heck, 2006]. The new version of the model (GHEISHA version 2002), that has been used in the present work,  incorporates certain modifications in kinematics through correction patches [Heck, 2006] thereby improving the energy and momentum conservation. 

The simulation package FLUKA describes particle interactions microscopically. It mainly employs resonance superposition models at low energies (up to $3$ to $5$ GeV), while relying on the two-string interaction model (the Dual Parton Model) [Capella et al., 1994] at intermediate energies. The basic conservation laws are obeyed at every single interaction level a priori. The resonance energies, widths, cross sections, and the branching ratios are extracted from data and the conservation laws by making explicit use of spin and isospin relations. In this model, the high energy hadron-nucleus interactions are described as a sequence of the Glauber–Gribov cascade, the generalized IntraNuclear cascade, pre-equilibrium emission and evaporation/fragmentation/fission and the final deexcitation, whereas, the nucleus-nucleus interactions above a few GeV/n are treated in FLUKA by interfacing with the DPMJET model [Ranft, 1995]. At low energies, a relativistic QMD model [Sorge, 1995] has been exploited here to describe the nucleus-nucleus interactions.   

In contrast, the UrQMD (Ultra-relativistic Quantum Molecular Dynamic) model [Bleicher et al., 1999] was originally designed  for simulating relativistic heavy ion collisions in the energy range from around $1$ AGeV to few hundred AGeV, that is the range of the (laboratory) energy in the case of the RHIC experiment. This microscopic model inherits the basic treatment of the baryonic equation of motion in the Quantum Molecular Dynamic model and describes the phenomenology of hadronic interactions at low and intermediate energies in terms of interactions between known hadrons and their resonances. The model does not use any parameterized cross sections. Instead, the projectile is allowed to hit a sufficiently large disk involving maximum collision parameters as a result of which the program consumes rather long a CPU time. We here add that both the UrQMD and the FLUKA describe fixed target data reasonably well. 

\section{Adapted Simulation procedure} 

The cosmic ray EAS simulation code CORSIKA is widely used in various studies ranging from TeV gamma-rays to the highest energy cosmic rays. Originally developed for the reproduction of extensive air showers with primary energies around the knee of the cosmic ray spectrum (i.e., around $10^{15}$ eV), CORSIKA has also found applications in the estimation of the flux of low energy secondary cosmic ray particles such as the neutrinos [Wentz et al. 2003]. In the present work, the intermediate-energy hadronic interaction models FLUKA, GHEISHA and the UrQMD have been used in combination with the high energy (above about $80 GeV/n$) hadronic interaction model QGSJET 01 version 1c [Kalmykov, Ostapchenko and Pavlov, 1997] in the framework of CORSIKA to generate secondary cosmic ray spectra. Due to the steeply falling energy spectrum of the primary cosmic rays, the contribution of the primary particles with energies above $80$ GeV/n on the secondary proton spectrum observed by the BESS-spectrometer is only about $15$ percent. The high energy hadronic interaction models are, therefore, not likely to have much effects on the low energy secondary proton spectrum.  

The fluxes of secondary particles obtained using CORSIKA have statistical as well as systematic errors. Besides the uncertainties in the theoretical treatment of the hadronic interactions, the main source of systematic errors in the calculation of secondary cosmic ray flux may, in fact, be the inaccuracies in the determination of absolute flux of the primary particles as was already mentioned in the section I above. Some details on the primary spectrum and specially, the effects of solar modulation and the geomagnetic field on such primary spectrum are, therefore, described in the following.   

\subsection{Primary spectra} 
In order to calculate the fluxes of the secondary cosmic rays, the primary cosmic ray energy spectrum is required as input in the simulations. The uncertainties in the determination of such primary flux have been substantially reduced in recent years with new precise measurements by the BESS (BESS-98 [Sanuki et al, 2000], the BESS-TeV [Haino et al., 2004]) and the AMS [Alcaraz et al.,2000a; J. Alcaraz et al. 2000b] experiments. The detectors of such experiments were calibrated by using the accelerator beams, thus ensuring the performance of those detectors. The spectra of primary cosmic ray protons observed by these experiments exhibit very good agreement with each other. In the case of the primary alpha particles, the agreement is not as good as that of proton spectra,  but still is quite reasonable. Since the proton is the dominant component, below $100$ GeV the difference in terms of nucleon flux is within just $4$ percent. For the reason stated before, in the present work the BESS primary spectra has been considered as input. 

BESS mission collected data on the absolute fluxes of primary protons and helium nuclei down to $1$ GeV [Sanuki et al, 2000]. At energies below about $10$ GeV the solar modulation effect is prominent. This effect introduces a time dependence on the absolute flux below this energy. The solar modulation can be handled by using the simple force field approximation [] according to which, the differential intensity $J(r,E,t)$ of cosmic ray species of total energy $E$ at a distance $r$ from the sun is given by

\begin{equation}
J(r,R,t) = \frac{E^{2}-m_{o}^{2}}{(E+\psi(t))^{2}-m_{o}^{2}} J[\infty, E+ \psi(t)], 
\end{equation}

where $m_{o}$ being the rest mass, $\psi(t) = (Ze/A) \phi(t)$ with $Z$ and $A$ being the mass number and atomic number of the cosmic ray species respectively and $\phi(t)$ is the modulation potential. $ J[\infty, E+ \psi(t)]$ describes the interstellar cosmic ray intensity. Depending on the solar activity the modulation potential can be obtained within theoretical models. Accordingly for each period of the solar cycle, the incident cosmic flux can be corrected for the different solar modulation effects using the above equation. 

Incorporating force field approximation in the Corsika program and considering the energy dependence of interstellar cosmic ray intensity as prescribed by Burger et al [Burger, 2000] i.e.  

\begin{equation}
J[\infty, E] = \frac{J_{o} P(E)^{-2.78} }{1+ 0.4866 P(E)^{-2.51}} \; , 
\end{equation}

where $J_{o}$ is the normalization factor and $P(E) = \sqrt{E^{2}-m_{o}^{2}}$, we first generated primary energy spectra (without considering the geomagnetic cutoff) for proton and helium with solar modulation potential $\phi = .565$ GeV [Usoskin et al., 2005] as applicable to BESS 98 flight and are compared with the BESS observation of primary proton and He spectra [ Sanuki et al, 2000] and the results are shown in Figures 1 and 2 respectively. 

\begin{figure}
\centering
\includegraphics[width=0.5\textwidth,clip]{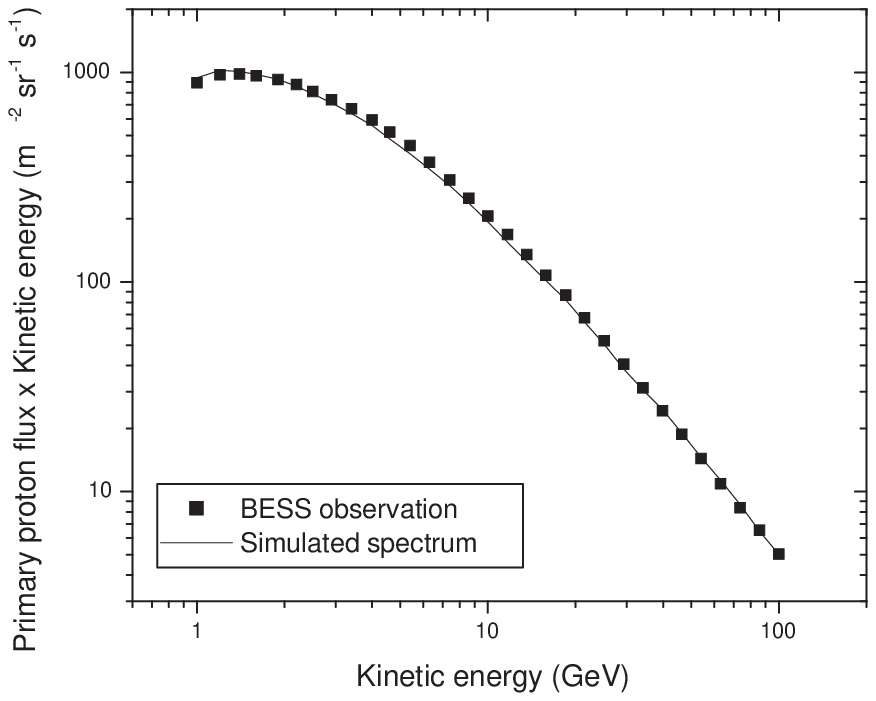} \hfill
\caption{Comparison of simulated primary spectra with the BESS 1998 observation.}
\end{figure}

\begin{figure}[h]
\centering
\includegraphics[width=0.45\textwidth,clip]{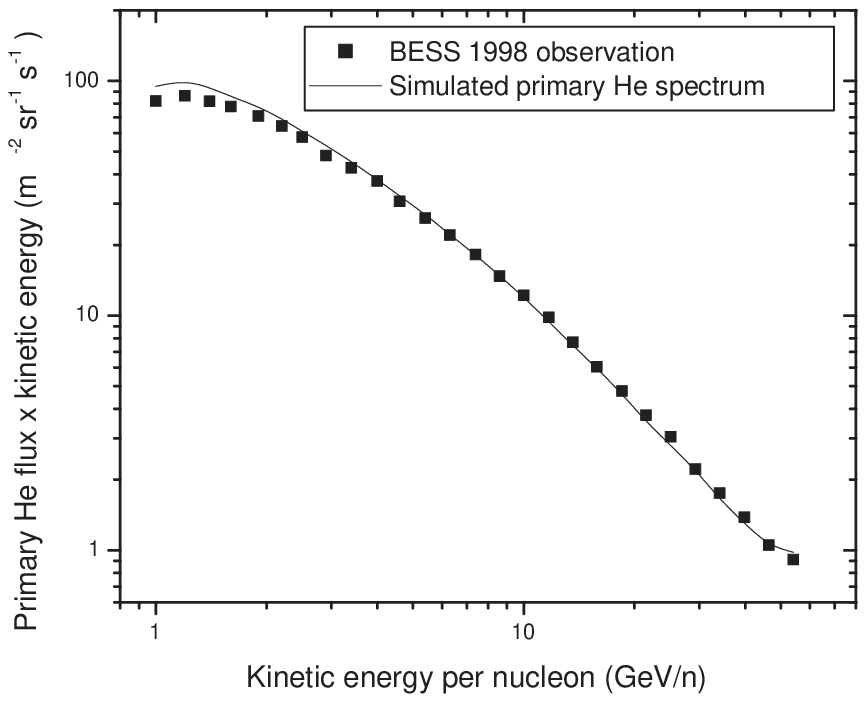} \hfill
\caption{Comparison of simulated primary spectra with the BESS 1998 observation.}
\end{figure}

It is clear from the figures that the agreement between simulated spectra and the observations is quite good. Subsequently, for the BESS high altitude measurement of secondary proton observation in September 1999 the primary cosmic ray spectra are generated taking solar modulation potential $\phi = .685$ GeV [Usoskin et al., 2005].    
The energy of the primary particle has been allowed to vary between the minimum of the geomagnetic cut off and $10^{14}$ eV.

\subsection{Geomagnetic cutoff}

The Earth's magnetic field imposes cutoff energy below which primary cosmic rays cannot penetrate to the atmosphere. This geomagnetic cutoff is an important ingredient of simulation aiming at estimation of secondary particle fluxes as it affects the primary flux on the location on the Earth. It also produces zenith and azimuth angle variation of cosmic ray flux at each position. Note that for measurement of primary cosmic ray spectra [] the BESS spectrometer was flown from Lynn Lake, Canada, where vertical geomagnetic cutoff is only about $0.5$ GeV. That is why no geomagnetic cutoff calculation has been performed for reproducing the BESS 1998 data as in Figures 1 and 2. However, the BESS experiment measured the secondary proton spectrum at Mt. Norikura, Japan where vertical cutoff is $11.2$ GeV. Thus precise estimation of geomagnetic cutoff is quite important in reproducing the secondary proton spectrum at Mt. Norikura.      

The cutoff needs to be computed for each primary cosmic ray particle. In the present work the geomagnetic cutoff calculations are performed using the (back) trajectory-tracing technique [Shea and  Smart, 1967]. The quiescent International Geomagnetic Reference Field (IGRF) model for 1995 [Sabaka et al., 1997] of the Earth's magnetic field has been used for the cutoff calculations. The cosmic ray rigidity cutoff, however, may not take an unique value, a penumbral region may exist in the particle rigidity range for a particular direction. In the penumbral region a complex series of allowed and forbidden cosmic ray trajectories coexist and an effective rigidity cutoff is obtained considering the transmission through the penumbra. 
The mean geomagnetic cutoff obtained from calculations for cosmic ray particles entering the atmosphere at the location of Mt. Norikura from different directions are shown in Figure 3. 

\begin{figure}[h]
\begin{center}
\includegraphics[width=0.45\textwidth,clip]{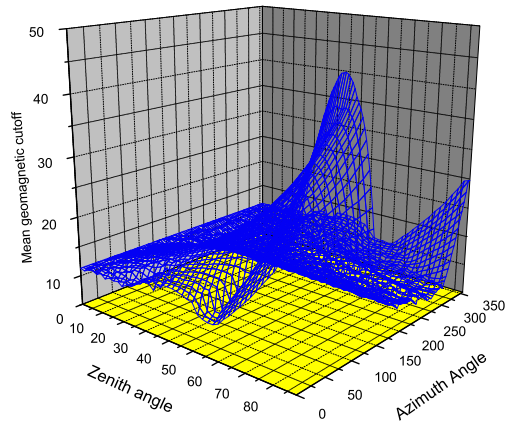} \hfill
\caption{Directional dependence of the mean geomagnetic cutoff for primary cosmic rays at the location of Mt. Norikura.}
\end{center}
\end{figure}

The penumbra region of the cutoff at Mt. Norikura location has also been determined and is shown in Figure 4. The width of the penumbra region at Mt. Norikura location is found to vary from $0$ GV to exceeding $4$ GV in some particular directions. 

\begin{figure}[h]
\begin{center}
\includegraphics[width=0.45\textwidth,clip]{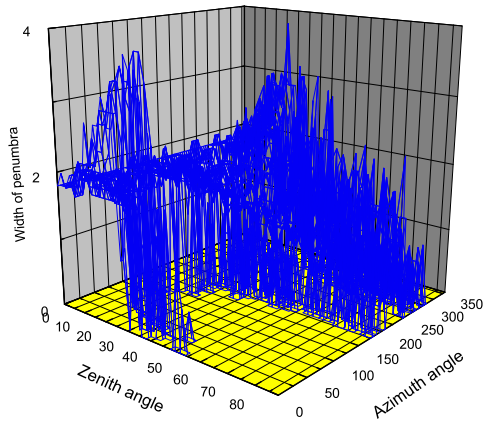} \hfill
\caption{The width of the geomagnetic penumbra region  at the location of Mt. Norikura.}
\end{center}
\end{figure}

The geomagnetic cutoff calculations have been used to modify the primary cosmic ray spectra. The charged particles bending due to Earth's magnetic field during shower development is taken care off by the CORSIKA program. 

\subsection{Other settings}

The fluxes of cosmic ray secondaries also depend on the atmospheric density profile. In the present work the US-standard atmospheric model with planar approximation has been considered. Such a model of atmosphere has been found reasonable in describing the atmospheric muon spectra at high altitude [Djemil et al., 2007]. The planar approximation works only when zenith angle of primary particles are below $70^{o}$. Since BESS observation of secondary cosmic rays was restricted to $cos \theta_{z} < 0.95$, planar approximation is sufficient to describe the data. Over and above the validity of this approximation for the present analysis has been ensured by generating data with the so called \lq curved \rq option for a particular set and comparing the results of two options for that set.     

Proton and helium are considered as primary particles. Instead of taking individually, primary nuclei heavier than helium are considered as groupwise, light, medium and heavy. Contributions of each variety of primaries to secondary proton/antiproton have been estimated separately and finally individual contributions are summed up to get the resultant secondary flux. Total nearly $2 \times 10^{7}$ events have been generated for estimation of the fluxes of secondary particles.   

\section{Simulated results and comparison with observation}
So far only a few measurements of cosmic-ray proton fluxes at mountain altitude were reported []. The most recent was due to the BESS group [Sanuki et al., 2003] by using a high resolution spectrometer with a large acceptance area consists of a superconducting magnet and a drift chamber based tracking system. They measured both atmospheric proton and antiproton spectra at Mt. Norikura, Japan ($742$ $g/cm^{2}$) in September, 1999. A threshold-type aerogel Cerenkov counter was used to distinguish protons and antiprotons from muons. Besides, the experiment was equipped with an electromagnetic shower counter to separate electrons and positrons from muons for muon analysis. The kinetic energy ranges covered are $0.25–3.3$ GeV/c. 
  
In Fig. 5, the total vertical proton spectrum as obtained from the simulation using all the three low energy interaction models at Mt. Norikura location are plotted and are compared with the observation of the BESS experiment. In the observation the zenith angle ($\theta_{z}$) of the measurements was limited as $Cos\theta_{z} \ge 0.95$. Hence the same restriction on the zenith angle has been imposed in deriving the simulated proton spectra. 

\begin{figure}[h]
\begin{center}
\includegraphics[width=0.45\textwidth,clip]{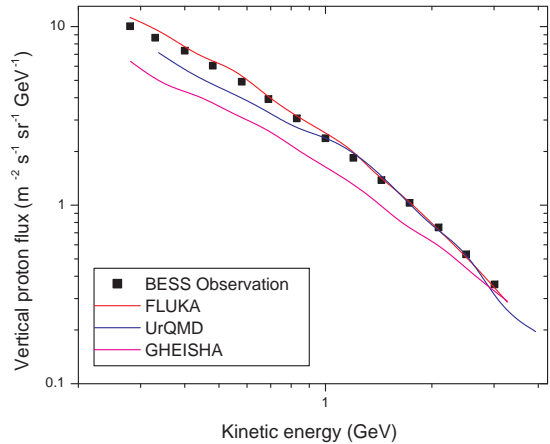} \hfill
\caption{The differential spectra of vertical proton at Mt. Norikura obtained for three low energy models. The results of BESS observation is also given for comparison.}
\end{center}
\end{figure}

It is clear from the Figure 5 that only the FLUKA describes the observed data properly over the whole energy range. The GHEISHA and the UrQMD give lower flux at low energy end than the BESS results. Particularly for GHEISHA, the simulated results are substantially lower than the BESS observations. The spectrum obtained using UrQMD and QGSJET combination shows reasonable agreement with the observations above secondary proton energy $1$ GeV but differs at lower energies. To ensure that the discrepancy at the low energy end originates from UrQMD but not from QGSJET, the secondary proton spectrum has also been generated using UrQMD and NEXUS combination and it has been found that similar deviation of the simulated data from the observation at low energy end persists for such a combination too. All the three models are found to approach to each other at higher energy range.

In Fig. 6, the comparison of the calculated zenith angle dependence of proton flux for two different kinetic energy range ($0.30 -0.36$ GeV and $1.90-2.29$ GeV) with the observed ones by the BESS experiment at Mt. Norikura has been shown. Here in calculating the flux the events with $Cos\theta_{z} \ge 0.85$ only have been selected to keep parity with the BESS observation. Though the simulated results contain some fluctuations, mainly of statistical origin, but it is evident from this plot that the nature of variation of proton flux with zenith angle for all the models is in accordance with the BESS observation. However, GHEISHA flux is found somewhat lower than the observation over the whole range of zenith angle, particularly at lower energy bin.  

\begin{figure}[h]
\begin{center}
\includegraphics[width=0.45\textwidth,clip]{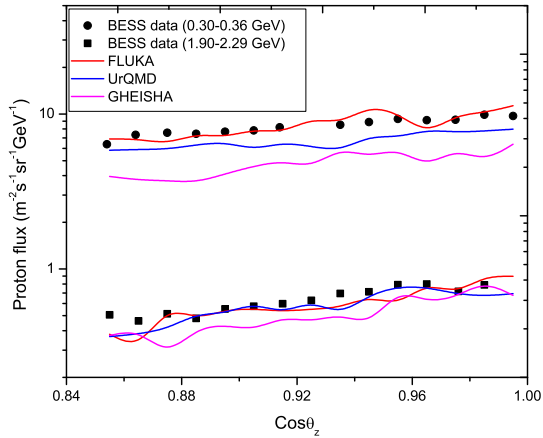}
\caption{The zenith angle dependence of proton flux at Mt. Norikura for two different kinetic energy bins obtained for three low energy models in comparison with the BESS observation.}
\end{center}
\end{figure}

The CR induced $\bar{p}$ flux in the atmosphere has been measured at mountain level by the BESS experiment. It is expected to be of purely atmospheric origin [Sanuki et al.,2003]. The simulated results for antiproton spectrum are compared with observations in Fig. 7. Surprisingly, in this case FLUKA could not reproduce the observe data well whereas antiproton spectrum due to UrQMD is found consistent with the experimentally observed results within the experimental error. GHEISHA too represents the observed data well though for this model simulated spectrum exhibits more fluctuations. Note that the simulation results for antiprotons are uncertain by around $10 \%$ at higher energy end and by nearly $30 \%$ at lower energy end due to limited statistics. 

\begin{figure}[h]
\begin{center}
\includegraphics[width=0.45\textwidth,clip]{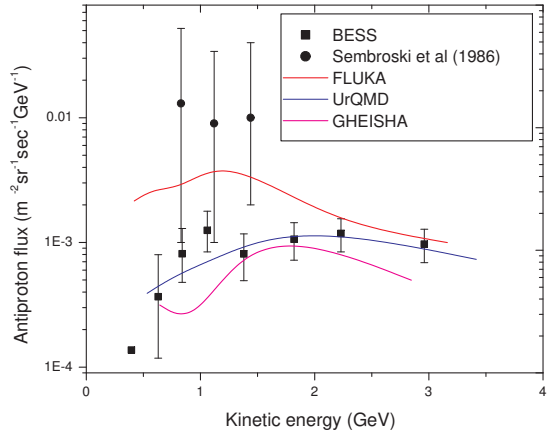}
\caption{The differential spectra of vertical antiproton at Mt. Norikura obtained for three low energy models. The results of BESS observation and those obtained by Sembroski et al. [1986] in a similar observational condition are also given for comparison.}
\end{center}
\end{figure}

\section{Conclusion}

We conclude the followings: 

1) The fluxes of cosmic ray secondary protons and antiprotons at mountain altitude are found quite sensitive to low energy interaction models.  

2) The present analysis suggests that among the three low energy models admissible in CORSIKA, FLUKA only produces secondary proton spectrum consistent with the BESS observation over the whole observed energy range. The model UrQMD works well at relatively higher energies whereas spectrum obtained with GHEISHA exhibits significant deviations from measured spectrum.

3)Atmospheric muon (as well as neutrino) spectrum is also known to susceptible on low energy interaction models [Battistoni, 2002]. In a recent study Djemil et al [2007] compared simulation results of atmospheric muon flux using low energy interaction models Fluka and UrQMD with the observations and noticed that the simulated muon flux for the two models reasonably differs at lower energies from each other but the muon flux due to both the models were found consistent with the experimentally observed results within the experimental error. The BESS observation of proton spectrum at mountain level is, however, clearly favor FLUKA over the UrQMD 1.3 model. Here a point to be noted that secondary protons arise from hadronic interactions in the forward kinematic region whereas muons come from the central region of interactions. Consequently secondary protons of same energy with muons are produced from primaries of comparatively lower energies. For instance muons with kinetic energy of $2$ GeV or more are produced predominantly by primary particles having energy at least $20$ to $50$ times higher whereas most protons of similar energy are produced by primary particles of energies less than $40$ GeV. Hence secondary protons not only probe the models in the forward kinematic range but also they are more sensitive to the models than muons of equal energies in the concerned energy region.  

4) Simulated antiproton spectrum with FLUKA shows significant disagreement with the BESS observation, fluxes obtained with FLUKA is found resonably higher than the observations though the difference between the simulated and the observed flux decreases at higher energies. However, FLUKA fluxes are found lower than the results obtained by Sembroski et al. [1986] in an experiment under similar observational condition ($747 g cm^{-2}$). It was earlier reported that the UrQMD 1.3 predictions on multiplicity of antiprotons in inelastic pp interactions overestimate the experimental data [Bleicher, 2000] in the SPS domain. Here the energy range is somewhat lower but the agreement of the model predictions with observation is quite well. 

For the given cosmic ray fluxes at the top of atmosphere, the antiproton flux at mountain altitude mainly relies on two factors: The inclusive antiproton production cross section in cosmic ray -air nuclei collisions and the propagation of antiprotons inside the atmosphere. 
The basic $\bar{p}$ production reaction is the inclusive $NN \rightarrow \bar{p}X$ process, N standing for the nucleon and X for any final hadronic state allowed in the process. The cross section data for the proton induced inclusive $\bar{p}$ production on nuclei and on the nucleon are not very accurate, particularly at energies around and below $1$ GeV because of
the lack of accelerator data over this energy range. Moreover, due to the high sensitivity to nuclear medium effects no reliable estimate of the low energy $\bar{p}$ production in cosmic ray- atmospheric nucleus collision can be made presently. The BESS results offer an indirect way to examine the antiproton production at low energies. The observed discrepancy between the FLUKA generated antiproton flux and the BESS observations as well as the large difference between the observational results obtained by BESS and Sembroski et al. [1986] suggest for further experimental and theoretical study on the topic.  
  
\section*{Acknowledgements} S.K. Ghosh, P.S. Joarder and Sibaji Raha thank the Department of Science and Technology (Govt. of India) for support under the IRHPA scheme.


\begin{thebibliography}{99}
\bibitem{ab:1} J. Alcaraz et al., Phys. Letts. B 472, 215 (2000a) 
\bibitem{ab:2} J. Alcaraz et al. Phys. Letts. B 494, 193 (2000b) 
\bibitem{ab:3} G. Battistoni, A. Ferrari, T. Montaruli and P. R. Sala, Astropart. Phys. (2002)
\bibitem{ab:4} R. A. Burger, M. S. Potgieter and B. Heber, J. Geophys. Res. 105, 447 (2000)
\bibitem{ab:5} M. Bleicher, M. Belkacem, S. A. Bass, S. Soff and H. Stoecker, Phys. Lett. B 485, 133 (2000); www.th.physik.uni-frankfurt.de/\~urqm/ 
\bibitem{ab:6} M. Bleicher et al., J. Phys. G: Nucl. Part. Phys. 25, 1859 (1999)
\bibitem{ab:7} J.N. Capdevielle et al., Report KfK 4998, Kernforschugszentrum Karlsruhe (1992)
\bibitem{ab:8} A. Capella, U. Sukhatme, C.-I. Tan and J. Tran Thanh Van, Phys. Rep. 236,225 (1994)
\bibitem{ab:9} T. Djemil, R. Attallah, and J. N. Capdevielle, J. Phys. G: Nucl. Part. Phys. 34, 2119 (2007)
\bibitem{ab:10} H.J. Drescher et al., Astropart. Phys. 21, 87 (2004)
\bibitem{ab:11} H.J. Drescher et al., Phys. Rep. 350, 93 (2001); www-subatech.in2p3.fr/\~theo/nexus
\bibitem{ab:12} R. Engel et al., Proc. 26th Int. Cosmic Ray Conf., Salt Lake City (USA), 1,  415 (1999)
\bibitem{ab:13} R.S. Fletcher et al. Phys. Rev. D50, 5710 (1994)
\bibitem{ab:14} A. Fass`o et al., FLUKA: Status and Prospective of Hadronic Applications, Proc. Monte Carlo 2000 Conf., Lisbon, Oct. 23-26, 2000; www.fluka.org/ 
\bibitem{ab:15} A. Ferrari and P. R. Sala, (1996) ATLAS int. note PHYS-No-086 (CERN, Geneva); 
\bibitem{ab:16} H. Fesefeldt, Report PITHA-85/02, RWTH Aachen (1985), 
\bibitem{ab:17} S. Haino et al., Phys. lett. B 594, 35 (2004) 
\bibitem{ab:18} D. Heck et al., Report FZKA 6019,Forschungszentrum Karlsruhe (1998); wwwik.fzk.de/˜heck/corsika/
\bibitem{ab:19} D. Heck, Nucl. Phys. Proc. Suppl. 151, 127 (2006)
\bibitem{ab:20} N.N. Kalmykov, S.S. Ostapchenko and A.I. Pavlov, Nucl. Phys. B (Proc. Suppl.) 52B, 17 (1997)
\bibitem{ab:21} S. Ostapchenko, Nucl. Phys. B (Proc. Suppl.) 151, 143 (2006)
\bibitem{ab:22} J. Ranft, Phys. Rev. D51, 64 (1995)
\bibitem{ab:23} T. J. Sabaka, R. L. Langel, and J. A. Conrad, J. Geomag. Geoelect., 49, 157 (1997)
\bibitem{ab:24} G. H. Sembroski et al., Phys. Rev. D33, 639 (1986).
\bibitem{ab:25} T. Sanuki et al., apj 545, 1135, (2000)
\bibitem{ab:26} T. Sanuki et al., Phys. Lett. B 577, 10 (2003)
\bibitem{ab:27} T. Sanuki et al., Phys. Rev. D (2007) 
\bibitem{ab:28} M. A. Shea, and D. F. Smart, J. Geophys. Res., 72, 2021 (1967)
\bibitem{ab:29} H. Sorge, Phys. Rev. C52, 3291 (1995)
\bibitem{ab:30} I. G. Usoskin et al., J. Geophys. Res. 110, A12108 (2005)
\bibitem{ab:31} J. Wentz et al., Phys. Rev. D67, 073020 (2003)
\bibitem{ab:32} K. Werner, Phys. Rep. 232, 87 (1993)

\end{thebibliography}
\end{document}